\documentclass[aps,pra,reprint,groupedaddress,showpacs]{revtex4-1}
\usepackage{graphicx}
\usepackage{amsmath}
\usepackage{amssymb}
\usepackage{amsfonts}
\usepackage{mathrsfs}
\usepackage{hyperref}

\newcommand{\ket}[1 ]{\mbox{$|#1\rangle$}}
\newcommand{\bra}[1]{\mbox{$\langle#1|$}}

\newcommand{\set}[1]{\{#1\}}
\newcommand{\outerp}[2]{\ket{#1}\bra{#2}}
\newcommand{\outerpp}[2]{\ket{#1}_p\bra{#2}}
\newcommand{\outerpc}[2]{\ket{#1}_c\bra{#2}}
\DeclareMathOperator{\tr}{tr}

\begin{document}

\title{Two-dimensional quantum walk under artificial magnetic field}
\author{\.{I}. Yal\c{c}{\i}nkaya}
\email[]{iyalcinkaya@sabanciuniv.edu}
\author{Z. Gedik}
\affiliation{Faculty of Engineering and Natural Sciences, Sabanc{\i} University, Tuzla 34956, \.{I}stanbul, Turkey}

\date{\today}

\begin{abstract}
  
We introduce the Peierls substitution to a two-dimensional discrete-time quantum walk on a square lattice to examine the spreading dynamics and the coin-position entanglement in the presence of an artificial gauge field. We use the ratio of the magnetic flux through the unit cell to the flux quantum as a control parameter. For a given flux ratio, we obtain faster spreading for a small number of steps and the walker tends to be highly localized around the origin. Moreover, the spreading of the walk can be suppressed and decreased within a limited time interval for specific rational values of flux ratio. When the flux ratio is an irrational number, even for a large number of steps, the spreading exhibit diffusive behavior rather than the well-known ballistic one as in the classical random walk and  there is a significant probability of finding the walker at the origin. We also analyze the coin-position entanglement and show that the asymptotic behavior vanishes when the flux ratio is different from zero and the coin-position entanglement become nearly maximal in a periodic manner in a long time range.  

\end{abstract}

\pacs{03.67.Ac,03.67.Bg,03.67.Lx,03.65.Vf,42.60.-v}
\maketitle

\section{\label{sec:int}Introduction}

A quantum computer was envisioned by Feynman as a device overcoming the difficulty of simulating quantum mechanical systems with classical computers \cite{feynman}. Today, ultracold atomic systems are of great interest for implementing highly controllable analogues of quantum systems under consideration \cite{bloch,buluta}. One way of creating and controlling such systems is trapping ultracold neutral atoms in periodic potentials of optical lattices formed by a laser. Since the atoms are neutral, external electric or magnetic fields, which are essential for quantum phenomena such as the quantum Hall effect and topological phases, have no effect on their trajectories. On the other hand, analogous effects can be implemented into optical lattices artificially to extend their simulation abilities. There are several proposals \cite{jaksch,dalibard} and experiments \cite{struck,lin,aidelsburger,garcia} focusing on the creation of tunable artificial gauge fields for ultracold neutral atoms in optical lattices by using atom-light interaction.

The discrete-time quantum walk (QW) was originally proposed by Aharonov \textit{et al.} as a quantum counterpart of the classical random walk, where the QW leads to ballistic spread of the walker rather than the diffusive one observed in the classical case \cite{aharonov}. Quantum walks are useful for developing new quantum algorithms \cite{ambainis} and they provide a model for universal quantum computation \cite{lovett}. They also supply a fertile framework for simulating other quantum systems such as topological phases \cite{kitagawa}, Anderson localization \cite{crespi,ghosh}, nonlinear $\delta$-kicked quantum systems \cite{buerschaper}, the breakdown of an electric-field driven system \cite{oka} and the creation of entanglement in bipartite systems \cite{schreiber}. Quantum walks can be realized experimentally in various physical systems including ultracold atoms in optical lattices \cite{perets,karski,zahringer,broome,peruzzo,sansoni,schreiber}. 

Recent experimental studies on optical lattices can allow the realization of QWs under artificial gauge fields. For example, it has been experimentally shown that the effect of an electric field on a charged particle can be mimicked by QWs in a one-dimensional optical lattice \cite{genske}. Also, a recent proposal by Boada \textit{et al.} utilizes photonic circuits for the realization of QWs under an artificial magnetic field \cite{boada}. In this paper we investigate the dynamics of two-dimensional (2D) QWs on a square lattice in the presence of an artificial magnetic field. For this purpose, we introduce position- and direction- dependent phases corresponding to the Peierls phases of the hopping terms between neighboring sites in the Hamiltonian representing a charged particle under a uniform magnetic field. We can control the propagation of the walker by changing the magnetic flux $\Phi$ through the unit cell. Depending on $\Phi$, we show that ballistic behavior can be suppressed within a time interval or can be completely broken.  It is known that, classical diffusive behavior in QWs is observed when quantum coherence is removed in some way, e.g., by decohering the coin and/or the position \cite{kendon}. We show that QW is also diffusive at long times if the magnetic flux ratio $\alpha=\Phi/\Phi_0$ ($\Phi_0$ being the flux quantum) is an irrational number and the walker remains highly localized at the origin throughout the walk. Moreover, we demonstrate that when $\alpha$ is chosen properly the walk stops to propagate and propagates back towards the origin during a limited time interval. We also analyze the entanglement between the coin and the position of the walker and show that the well-known asymptotic behavior vanishes when $\alpha \neq 0$. We observe that the coin and the position become maximally entangled at specific steps under the effect of the artificial magnetic field on a long time scale.

This article is organized as follows. In Sec. \ref{sec:def} we give a brief overview of QWs and introduce the Peierls model to the formalism. In Sec. \ref{sec:rif} we compare the behavior of the QW under rational and irrational $\alpha$'s and we demonstrate the localization of the walker. In Sec. \ref{sec:ent} we examine the effect of the magnetic field on coin-position entanglement. In Sec. \ref{sec:con} we summarize our results.

\section{\label{sec:def}Peierls model in QWs}

In analogy to the classical random walk, the master equation for the discrete-time quantum walk is given by $\ket{\Psi_{t+1}}=\mathcal{U}\ket{\Psi_{t}}$, where $\mathcal{U}$ is a unitary transformation describing the time evolution of the state vector $\ket{\Psi_{t}}$ in discrete bipartite coin-position Hilbert space $\mathscr{H}_c \otimes \mathscr{H}_p$  spanned by $\set{\ket{0}_c,\ket{1}_c}$ and $\set{\ket{n}_p\ |\ n \in \mathbb{Z}}$, respectively. The operator $\mathcal{U}=\mathcal{S}(\mathcal{C}\otimes\mathcal{I})$ is called a step of the walk and it is composed of a shift operator $\mathcal{S}$ and a coin operator $\mathcal{C}$. The coin operator acts only on the coin space and it can be any unitary operation in SU(2). In the first proposal of the QW \cite{aharonov}, $\mathcal{C}$ was chosen as the Hadamard gate

\begin{equation}
\mathcal{C}_H=\frac{1}{\sqrt{2}}(\outerpc{0}{0}+\outerpc{0}{1}+\outerpc{1}{0}-\outerpc{1}{1}),
\label{eq:hd}
\end{equation}

\noindent
which is the one we use throughout the paper. The conditional shift operator on a line is

\begin{equation}
\mathcal{S}=\sum_{n=-\infty}^{+\infty}(\outerpc{0}{0}\otimes\outerpp{n-1}{n} + \outerpc{1}{1}\otimes\outerpp{n+1}{n}),
\end{equation}

\noindent
which acts on both spaces (coin and position) and it moves the walker to the left (right) when the coin component is in the state $\ket{0}_c$ ($\ket{1}_c$). If the walk starts with the initial state $\ket{\Psi_{0}}$, after $t$ steps, the final state becomes

\begin{equation}
\ket{\Psi_t}=\mathcal{U}^t\ket{\Psi_{0}}=\sum_{c,n}a_{n,c;t}\ket{c,n}
\end{equation}

\noindent
where $a_{n,c;t}$ are the site amplitudes and $c \in \set{0,1}$. The probability of being found at any position $P_{n;t}=\sum_c |a_{n,c;t}|^2 $ is calculated by summing over the probabilities in the coin space.

Quantum walks on a line can be extended to higher dimensions by enlarging the coin and the position spaces. For 2D QWs, the coin space can be chosen as four dimensional with the basis $\set{\ket{0}_c,\ket{1}_c,\ket{2}_c,\ket{3}_c}$, while the position space is spanned by $\set{\ket{n,m}_p\ |\ n,m \in  \mathbb{Z}}$. The coin can be interpreted as a single four-level coin or two different two-level coins. For both cases, the coin operator can be factorizable, i.e., $\mathcal{C}=\mathcal{C}_H \otimes \mathcal{C}_H$, or non-factorizable, e.g., the Grover coin (the so-called Grover walk) or the discrete Fourier transform coin \cite{tregenna}. When the walker component $\ket{n,m}_p$ corresponds to the position eigenstates of the walker on a square lattice, the shift operator is defined as a single operation that moves the walker in the left down, left up, right down and right up directions (i.e., towards corners) for respective coin states \cite{mackay}. On the other hand, experimentally, the walker can step to the nearest neighbors rather than the corners. Therefore, if we want the walker to be found at the corners in one step, we have to implement the shift operator as two separate operations, i.e., a shift along one axis followed by a second shift along the other axis. An alternative method for realization of 2D QWs is to use a single two-level coin instead of a four-level coin \cite{ambainis:soda2005,franco}. In this scheme, a step is defined as $\mathcal{U}=\mathcal{S}_y(\mathcal{C}\otimes\mathcal{I})\mathcal{S}_x(\mathcal{C}\otimes\mathcal{I})$, where the walker is first shifted along the $x$ direction followed by a shift along $y$ direction. The Hadamard operator in Eq. (\ref{eq:hd}) is chosen as the coin operator and it is applied before each shift. It has been shown that, with this scheme, the probability distribution of the Grover walk can be mimicked. This alternate scheme has advantages over a walk with a four-level coin when experimental aspects for square lattices are considered. Therefore, in this paper, we use this alternate scheme.

\begin{figure}
\includegraphics[scale=1]{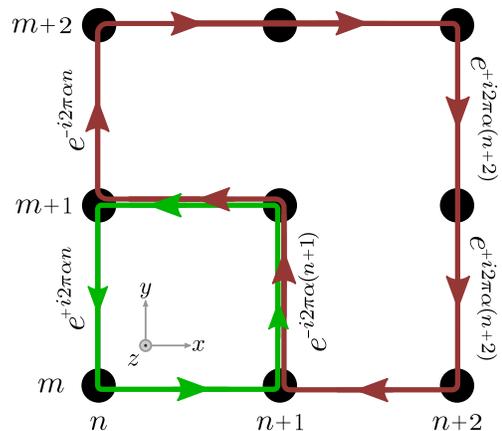}
\caption{Phases gained by the walker while hopping between neighboring sites in the $y$ direction. These phases are analogous to those acquired by a charged particle subjected to a constant external magnetic field in the $z$ direction, in the Landau gauge. Black circles denote the sites of the square lattice labeled by the integer pairs $(n,m)$. The arrows show two possible trajectories. The walker gains a total phase of $e^{-i2\pi\alpha}$ along the smaller paths and $e^{-i2\pi(3\alpha)}$ along the larger paths. In general, for an arbitrary closed path, the coefficient of $\alpha$  gives the number of unit cells enclosed. No phase is gained along the $x$ direction.}
\label{fig:pp}
\end{figure}

\begin{figure*}[t]
\centering
\begin{tabular}{cc}
(a) & (b) \\
\multicolumn{2}{c}{\includegraphics[scale=0.89]{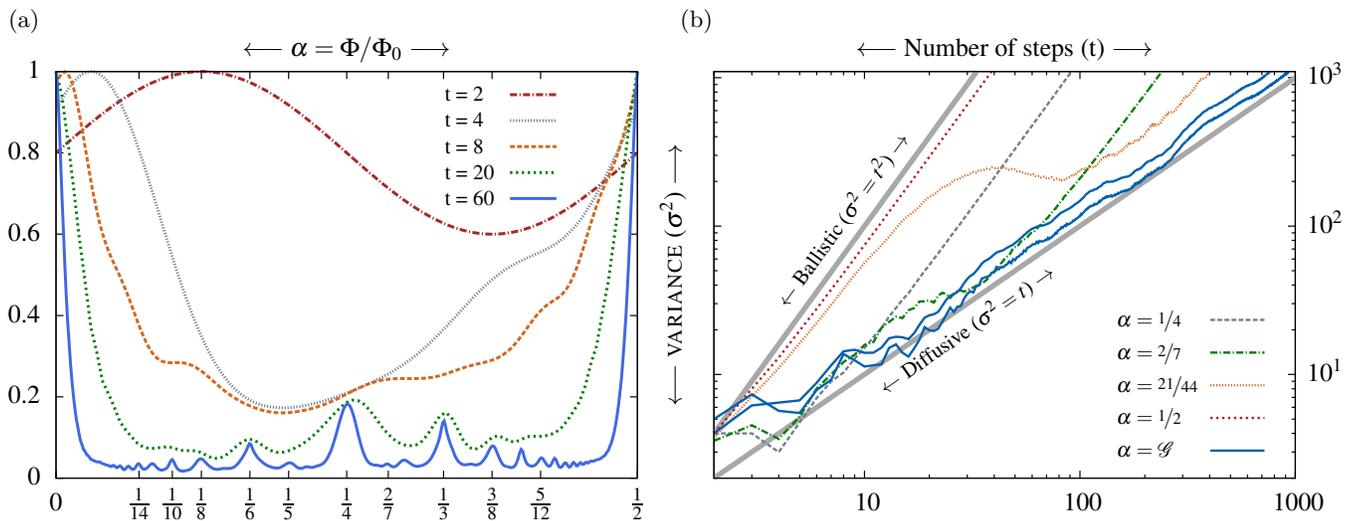}}
\end{tabular}
\caption{(a) Spreading of the walk after a different number of steps $t$ with respect to the magnetic flux $\Phi$ in units of flux quantum $\Phi_0$. The symmetrical initial state $\ket{\Psi_0^s}$ is used. Only one period of variances is drawn within the interval $[0,1]$. The vertical axis is rescaled with the maximum value of the variance for each $t$. For the rational values of $\alpha$, sharp peaks become apparent when the number of steps is increased. (b) Spreading of the walk with respect to the number of steps for different values of $\alpha$. When $\alpha$ is the golden ratio $\mathscr{G}$, we used two different initial conditions. The lower solid line corresponds to the symmetric case used for the other $\alpha$ values and the upper one is $\ket{\Psi_0}=\ket{0}_c\otimes\ket{0,0}_p$.}
\label{fig:sqw}
\end{figure*}

We consider QWs on a square optical lattice with a tunable artificial gauge field. We label the space coordinates as $x=na$ and $y=ma$, where $a$ is the lattice constant. We choose the symmetrical initial state $\ket{\Psi_0^s}=\ket{\psi}_c \otimes \ket{0,0}_p$, where both $n$ and $m$ are defined as zero and $\ket{\psi}_c=\frac{1}{\sqrt{2}}(\ket{0}_c+i\ket{1}_c)$. We introduce the shift operator in the $y$ direction as

\begin{equation}
\begin{split}
\mathcal{S}_y = &\outerpc{0}{0}\otimes \sum_{n,m} e^{+i2\pi\alpha n} \outerpp{n,m-1}{n,m} \\
+ &\outerpc{1}{1}\otimes \sum_{n,m} e^{-i2\pi\alpha n} \outerpp{n,m+1}{n,m}
\end{split}
\label{eq:sy}
\end{equation}

\noindent
where $\alpha \in [0,1]$ is the tuning parameter and $e^{\pm i 2\pi\alpha n}$ are both site and direction-dependent phases of hopping terms between neighboring lattice sites. At this point our approach differs from \cite{genske} and \cite{cedzich}, where only site-dependent phases on a one-dimensional lattice are used. When we consider the motion of a charged particle on a square lattice under a uniform magnetic field $\mathbf{B}=B_0\hat{\mathbf{z}}$, the appropriate Peierls substitution is given by \cite{peierls,hofstadter}

\begin{equation}
t \longrightarrow t \exp{\left(-i\frac{2\pi}{\Phi_0}\int_{\mathbf{r_1}}^{\mathbf{r_2}} \mathbf{A}.d\mathbf{l}\right)}
\label{eq:pp}
\end{equation}

\noindent
where $t$ is the nearest-neighbor hopping amplitude and $\Phi_0=h/e \approx 4.14\times10^{-15}\ \text{Wb}$ is the flux quantum ($h$ and $e$ being the Planck constant and elementary charge, respectively). Here $\mathbf{r_1}$ and $\mathbf{r_2}$ denote the initial and  final positions of the particle, respectively, and the integral is evaluated along the line connecting these points. If the vector potential is chosen as $\mathbf{A}=B_0x\hat{\mathbf{y}}$ (Landau gauge), the transition amplitude along the $x$ direction remains unaffected while along the $y$ direction it gains a factor of

\begin{equation}
\exp{\left(\pm i 2\pi\frac{\Phi}{\Phi_0}n\right)}
\end{equation}

\noindent
when hopping from site $m$ to $m\pm 1$. Here $\Phi=B_0 a^2$ is the magnetic flux through the unit cell. Therefore, we interpret the phases in Eq. (\ref{eq:sy}) as artificial Peierls phases shown in Fig. \ref{fig:pp}. In this paper, we consider both rational and irrational flux ratios $\alpha=\Phi/\Phi_0$ to find out the effects of the magnetic field on the characteristics of QWs such as the variance, participation ratio, and coin-position entanglement. 

\begin{figure*}[t]
\centering
\begin{tabular}{cc}
(a) & (b) \\
\multicolumn{2}{c}{\includegraphics[scale=0.358]{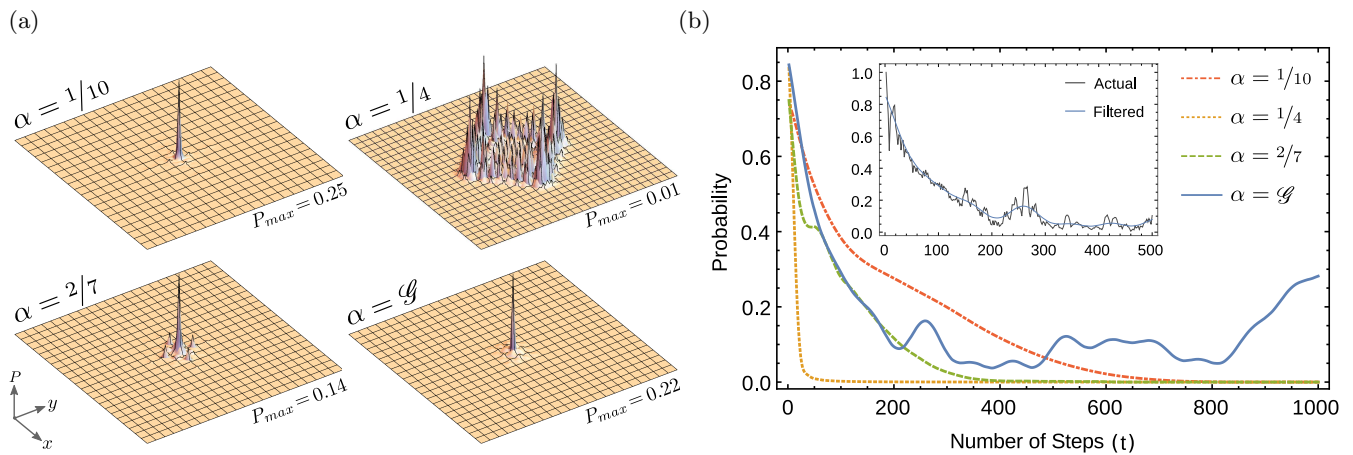}}
\end{tabular}
\caption{(a) Probability distribution of the walk after $100$ steps for different flux ratios. Here $\mathscr{G}$ is the golden ratio. (b) Probability of finding the walker in the vicinity of the origin after $t$ steps: the sum of the probabilities at sites $(n,m)$, where $n,m \in \set{-2,0,2}$ with respect to the number of steps. Since an odd number of steps gives vanishing probability, the probabilities for only an even number of steps are shown. Data are smoothed out via Gaussian filtering for simplicity. The inset shows a comparison of the actual data with the filtered data for $\alpha = \mathscr{G}$.}
\label{fig:probori}
\end{figure*}

\section{\label{sec:rif}Rational vs Irrational Flux Ratios}

We consider the variance $\sigma^2_t$ of the probability distribution, 

\begin{equation}
\sigma^2_t= \sum_{n,m}(n^2+m^2)  P_{(n,m);t},
\end{equation}

\noindent
as a measure of the spreading. Here, $P_{(n,m);t}$ is the probability of the walker being found at site $(n,m)$ after $t$ steps and we look for how the variance behaves under rational and irrational magnetic fields. When the flux ratio $\alpha$ is an integer, the original translational symmetry of the lattice is preserved. Similarly, when it is a rational number such as $p/q$, where both $p$ and $q$ are coprime integers, translational symmetry is preserved only if the unit cell is considered $q$ times as large. However, for irrational $\alpha$ values, the number of unit cells enclosed by the walker (see Fig. \ref{fig:pp}) is incommensurable with the parameter $\alpha$. Therefore, we cannot exploit a rescaling as we did in the rational case.  In Fig. \ref{fig:sqw}(a), we show the change of the variance with respect to $\alpha$ for different step numbers. Although the variance is meaningful for the walks with a large number of steps, we intentionally present the results for a small number of steps to demonstrate an interesting effect of magnetic fields on QWs. Note that since the variance repeats itself with a period of $1/2$ within the interval $[0,1]$, only one period is drawn for each case. When $t=2,4$, or $8$, for each case there are two maxima at non-zero $\alpha$ values. While the number of steps is increasing, the maxima move to the left and their positions converge to $\alpha=0$ and $\alpha=1/2$. For example, the analytic expression of the variance for two steps is

\begin{equation}
\sigma^2_2=3+2\cos^2(2\pi\alpha-\frac{\pi}{4}).
\label{eq:avar}
\end{equation}

\noindent
It is clear that the maximum value of Eq. (\ref{eq:avar}) occurs at $\alpha=1/8$ and $5/8$. It is also notable that in this case the probability at the origin, 

\begin{equation}
P_{(0,0);2}=\frac{1}{2}\cos^2(2\pi\alpha+\frac{\pi}{4}),
\end{equation}

\noindent
becomes zero because of the destructive interference of the incoming amplitudes. In other words, the walker avoids stepping into the center at the second step as a result of the applied field. Similarly, maxima for four and eight steps occur at $\alpha=3/100$ and $\alpha=3/400$, respectively. Hence, with an appropriate choice of the $\alpha$, the walk spreads faster than the case where there is no applied field for a small number of steps. For $20$ and $60$ steps, spreading of the walk is dramatically reduced by the non zero values of $\alpha$ except $1/2$. However, some peaks occur at the rational values of $\alpha$ and more peaks become apparent when the number of steps is larger. This happens because the more steps taken, the more routes that satisfy translational symmetry corresponding to different $\alpha$ values can be followed. Translational symmetry due to any rational number $\alpha$ requires the walker to take at least a number of steps that is enough to follow a closed path covering the unit cell rescaled by $\alpha$ appropriately. Therefore, we increase the $\alpha$ resolution of the walk by increasing the step number and thus the walk becomes more sensitive to the $\alpha$ values. By counting the number of peaks in Fig. \ref{fig:sqw}(a) roughly, we see that while the system is able to resolve 4 of those symmetries at 20 steps, it can resolve more than 20 if we increase the step number to 60. 

\begin{figure*}[t]
\centering
\begin{tabular}{ll}
(a) & (b) \\
\includegraphics[width=0.49\textwidth]{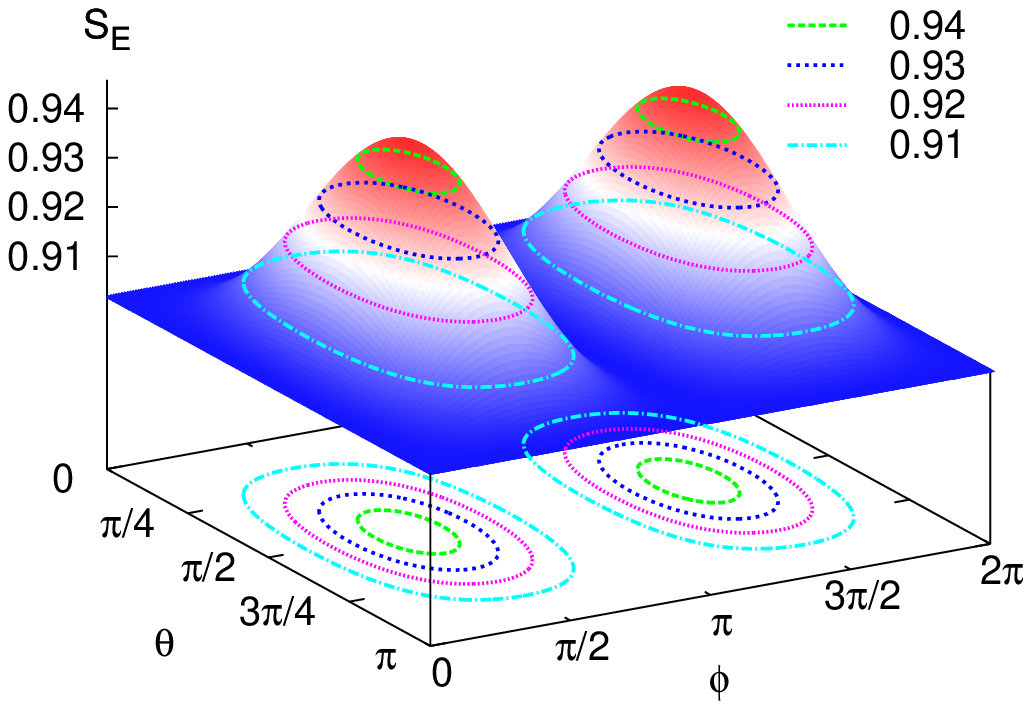} &  
\includegraphics[width=0.49\textwidth]{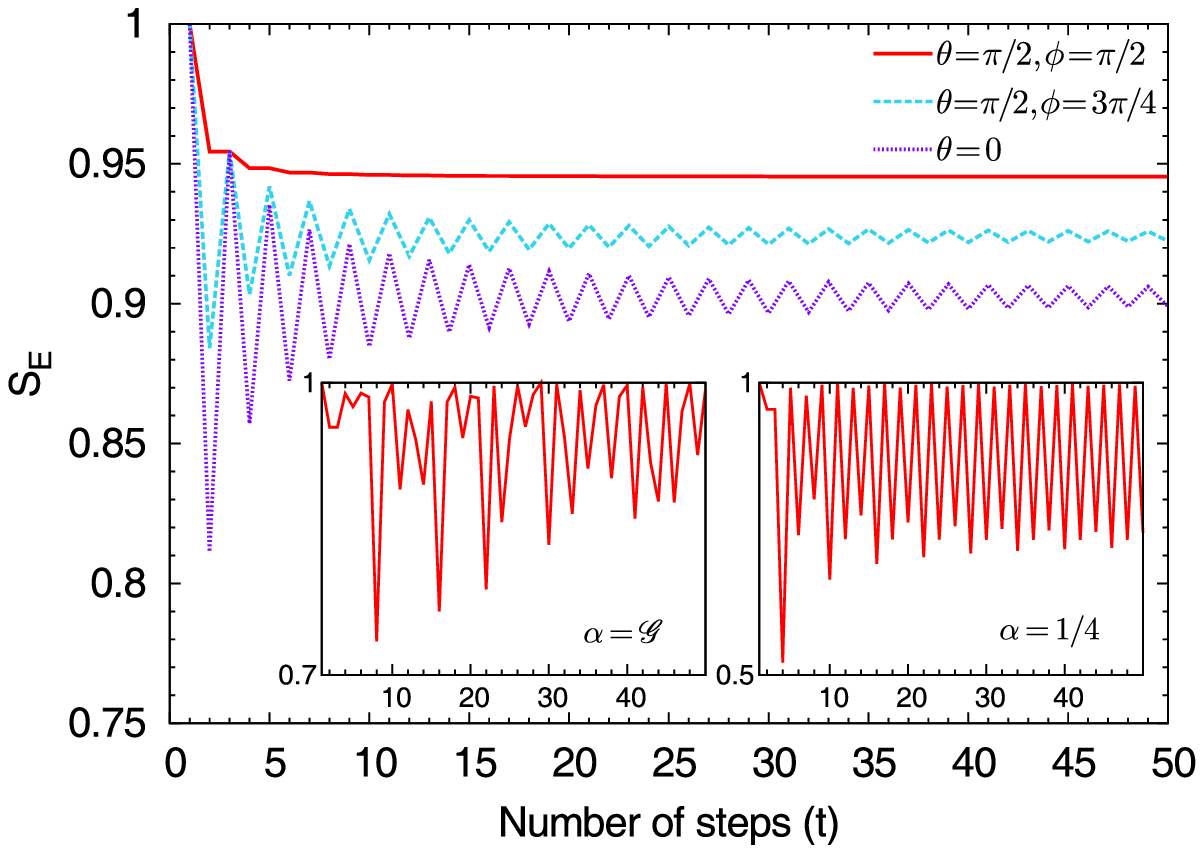}
\end{tabular}
\caption{(a) Asymptotic values of coin-position entanglement of a QW on a square lattice with two-level initial coin states $\ket{\psi}_c=\cos{\frac{\theta}{2}}\ket{0}+e^{i\phi}\sin{\frac{\theta}{2}}\ket{1}$ for $\alpha=0$. The figure is symmetrical about $\phi=\pi$ and varies between its maximum $\simeq 0.903$ and minimum $\simeq 0.945$. The maximum value is attained for symmetrical initial coin states $(\theta,\phi)=(\pi/2,\pi/2)$ and $(\pi/2,3\pi/2)$. (b) Time dependence of the entanglement for specific values of the initial coin state. The asymptotic behavior can be observed explicitly. The insets show that when $\alpha \neq0$, the asymptotic behavior vanishes. For $\alpha = \mathscr{G}$ (left inset) and for $\alpha = 1/4$ (right inset), after the fourth step, the coin and the position become almost maximally entangled ($S_E\simeq0.99$) every two steps.}
\label{fig:ent}
\end{figure*}

In Fig. \ref{fig:sqw}(a) we see that if $\alpha=1/2$, the variance is the same as the case where there is no magnetic flux $\alpha=0$ for a large number of steps. Actually, the walk has exactly the same dynamics for this two case and the same probability distribution is obtained after each step. The reason is that in the second and each succeeding step, translational symmetry is preserved, i.e., in each step the gained phases cancel each other after following a closed path that covers two unit cells. Hence, there is no net effect of magnetic flux $\alpha = 1/2$ on the walk. Note that no components of the wave function interfere with each other at the first step, i.e., the wave function only spreads towards the nearest corners. Therefore, the possible effects of the magnetic field occur after the first step for each walk under our consideration.

In Fig. \ref{fig:sqw}(b) we demonstrate that, for two different initial states, the walk spreads diffusively in a long time range if $\alpha$ is an irrational number and the golden ratio ($\mathscr{G}=\frac{\sqrt{5}-1}{2}$), due to the broken translational symmetry of the lattice. 
In this case, there is no rescaled unit cell that is commensurable with the original one to obtain a translational symmetry. In contrast, if $\alpha$ is rational, after few steps, spreading of the walk returns to its usual ballistic behavior. A remarkable case in which the spreading is nearly stopped and reversed during steps $35$-$50$ and  $50$-$85$, respectively, occurs for $\alpha=21/44$. Also, further simulations show that similar effects as in  $\alpha=21/44$ can be observed if $\alpha$ is chosen sufficiently close to $0$ or $1/2$. In general, we can conjecture that when $\alpha$ is a rational number, even if the spreading of the walk fluctuates at the beginning,  it will revert to its original ballistic behavior after a finite number of steps. Only irrational $\alpha$ values permanently suppress the spreading and result in a diffusive behavior.

Figure \ref{fig:probori}(a) demonstrates that whether $\alpha$ is a rational or an irrational number, the walker tends to be localized around the origin. However, some values of $\alpha$ cause stronger localization. In Fig. \ref{fig:probori}(b) the sum of probabilities of the walker being found at the origin and some nearby sites with respect to the number of steps is given. When $\alpha$ is a rational number, even though the robustness of the localization against step number changes for different $\alpha$'s, the probability of finding the particle around the origin converges to zero while the number of steps is increasing. In contrast,  when $\alpha = \mathscr{G}$, the probability does not converge to zero within the interval $[0,1000]$ and moreover it increases to approximately $0.3$ when $t=1000$. Although we examine the sum of probabilities at several sites around the origin, Fig. \ref{fig:probori}(a) ensures that the significant contribution to this sum comes from the origin. Therefore, an irrational flux ratio guarantees that the probability is the highest at the origin in the range up to $1000$ steps.

Although our results include only one irrational number (the golden ratio), further simulations for $1/\pi$, $1/e$, $1/\sqrt{2}$, and $1/\zeta (3)$ show that all of them exhibit diffusive behavior on average. As we have mentioned above, the reason is the incommensurability of the number of unit cells enclosed by the walker and the magnetic flux ratio for each case. Note that, as shown in Fig. \ref{fig:sqw}(b), the walk can exhibit higher spreading rates temporarily, e.g., between $t=250$ and $600$ for $\alpha = \mathscr{G}$, but on average, spreading fluctuates around the diffusive trend. However, these temporary deviations from the diffusive spreading can extend over relatively large time intervals for some irrational numbers. Among our simulations, the only example is $\alpha=1/\pi$, where we observe such a deviation over the interval from $t=100$ to $2000$. A possible explanation of this behavior can be given by answering the question of how well an irrational number can be approximated by the rationals. It is known that the best rational approximations to an irrational number are found in its convergents of continued fraction which are represented by $c_i=[a_0,a_1,\cdots,a_i]=p_i/q_i$ \cite{rockett}. By adding more terms, we obtain better approximations. Since the Peierls phases can be written as $e^{i2\pi c_i}e^{\pm i2\pi \alpha_{\text{err},i}}$ where $\alpha_{\text{err},i}=|c_i-\alpha|$, the phases  $e^{i2\pi c_i}$ allow various translational symmetries determined by $q_i$. We also know that if $\alpha_{\text{err},i}$ is sufficiently close to $0$ or $1/2$ , the effects result from $e^{\pm i2\pi \alpha_{\text{err},i}}$ cannot be observed for a while (see $\alpha=21/44$ in Fig. \ref{fig:sqw} and $\alpha=1/500$ in Fig. \ref{fig:ent2}). While the number of steps increases, the walk continuously tries to catch different translational symmetries required by the convergents but due to the accumulation of $e^{\pm i2\pi \alpha_{\text{err},i}}$ phases, this can never be done completely, which thereby results in temporary higher spreading rates. Therefore, if the $\alpha_{\text{err}}$'s are sufficiently small for a given irrational $\alpha$, we can expect to observe higher spreading rates extending over relatively long time intervals. What makes $1/\pi$ special is that its  convergents have the minimal errors when we compare it with the other irrational numbers we used. It has the best rational approximations among the others and this can be the reason for the deviations in long ranges. A detailed analysis including a careful comparison of the convergents of different irrational numbers and the corresponding errors is left to future research.

\section{\label{sec:ent}Effect of magnetic fields on Entanglement}

The shift operator generates entanglement between the coin and the position degrees of freedom \cite{carneiro,abal,annabestani}. Since the step operator $\mathcal{U}$ is unitary, the density matrix $\rho_t=\outerp{\Psi_t}{\Psi_t}$ at any step $t$ will be a pure state. Therefore, it is convenient to use the von Neumann entropy as a measure of the coin-position entanglement

\begin{equation}
S_E=-\sum_i{\lambda_i}\log_2{\lambda_i},
\end{equation}

\noindent
where the $\lambda_i$'s are the eigenvalues of the reduced density matrix $\rho_c=\tr_p{(\rho_t)}$ obtained by tracing out the position degree of freedom. The asymptotic behavior of the entanglement in QWs on a line and 2D QWs with a four-level coin is already known \cite{carneiro,abal}. Here we demonstrate that a 2D QW  with a two-level coin also exhibits the same behavior for all initial coin states for $\alpha=0$.  In Fig. \ref{fig:ent}(a) the dependence of the asymptotic values of coin-position entanglement on initial coin states is given. It can be seen that entanglement varies between a maximum of $0.945$ and a minimum of $0.903$. The maximum value occurs for the symmetrical initial state $\ket{\Psi_0^s}$. In Fig. \ref{fig:ent}(b) time dependence of coin-position entanglement for three specific initial coin states is given. For all cases, entanglement decays to an asymptotic value. 

\begin{figure}[t]
\centering
\includegraphics[width=0.49\textwidth]{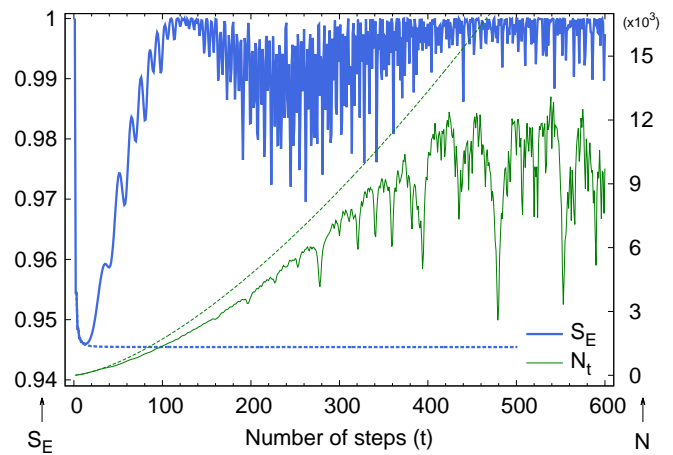}
\caption{\label{fig:ent2} Comparison between coin-position entanglement (thick) $S_E$ and participation ratio (thin) $N$ when $\alpha=1/500$. Dashed lines represent the behavior of each case when $\alpha =0$.}
\end{figure}

The inset in Fig. \ref{fig:ent}(b) shows the change in coin-position entanglement in the presence of a magnetic field (for $\alpha\neq 0$ and $\alpha\neq 1/2$). For both $\alpha=1/4$ and $\alpha=\mathscr{G}$, the asymptotic behavior disappears completely. Although we show only two cases for simplicity, we observe similar behaviors for the other $\alpha$ values. When $\alpha=1/4$, the value $S_E \simeq 0.99$ indicates that the coin and the position are almost maximally entangled in every odd step after the fourth step. When $\alpha=\mathscr{G}$, we obtain almost maximally entangled states in a quasi-periodic manner in a long time range again. To gain insight into the cause of the large values of entanglement, we consider the participation ratio at step number $t$,
\begin{equation}
N=\left( \sum_{n,m}{P_{n,m;t}^2}\right)^{-1},
\end{equation}

\noindent
which can be interpreted as an estimate for the number of sites over which the walker is distributed. While $N=1$ indicates that the walker is completely localized at only one site, $N=d$ indicates a uniform spreading over $d$ sites. In Fig. \ref{fig:ent2} we compare the coin-position entanglement and the participation ratio for a small magnetic field. During the first $10$ steps, we see that the entanglement is exactly the same for $\alpha=0$ and $\alpha=1/500$. Further simulations for smaller values of $\alpha$ show that the entanglement is the same as the $\alpha=0$ case even for a larger number of steps. This observation suggests that the effects of the magnetic field start to appear after a sufficient amount of phases has accumulated. After about $400$ steps, the participation ratio starts to oscillate strongly and we also observe quasi periodic oscillations in $S_E$, which results in almost maximally entangled states as in the insets of Fig. \ref{fig:ent}(b). Since strong oscillations in the participation ratio imply the tendency of the walker to return to the origin, we can conclude that localization of the walker plays an important role in the large values of coin-position entanglement in the presence of a magnetic field.

\section{\label{sec:con}Conclusion}

We have studied the spreading properties and coin-position entanglement for the QW on a square lattice under an artificial magnetic field. We have demonstrated that the presence of such fields increases the spreading for a small number of steps. Moreover, irrational flux ratios cause a diffusive spreading rather than a ballistic one even in the long time range because of the broken translational symmetry. This  also causes the probability at the origin to be the highest even for a large number of steps. For rational flux ratios, ballistic spreading can be suppressed within a limited time range. However, the walk returns to the original ballistic behavior after a finite number of steps. 

We have demonstrated that the coin-position entanglement in a QW on a square lattice with a single two-level coin exhibits an asymptotic behavior as in the four-level case. We have also shown that this behavior changes in the presence of an artificial gauge field and it is possible to keep the coin and the position maximally entangled in a long time range if the field is chosen appropriately. However, we have not observed any distinguishing property of irrational or rational magnetic flux ratios while examining the coin-position entanglement.

Finally, we note that our scheme may be realized with today's technology and some of our results, especially the ones that require only a few number of steps, may be verified. We believe that our work can provide a further step towards simulating many-body quantum systems in gauge fields by engineering the interactions of ultracold atoms with light.

\begin{acknowledgements}

We would like to thank O. Benli and G. Karpat for helpful discussions.

\end{acknowledgements}

\bibliography{bibliography}

\end{document}